\def\bi{\bigskip}
\def\bq{\begin{eqnarray}}
\def\eq{\end{eqnarray}}
\def\noi{\noindent}

\def\ct{\centerline}

\magnification=1200

\baselineskip 20 pt
\vglue 2 cm
\ct {\bf Inclusive $J/\psi$ production in $\Upsilon$ decay via  color-octet 
mechanisms.\footnote*{{\rm Work supported in part by Universidad de 
Guanajuato, Mexico, in part by Conacyt-Mexico, under Grant 933053 and 
Project 3979PE-9608.}}}
\vglue 2 cm  
\ct {\it M. Napsuciale \footnote{**}{{\rm E-mail: mauro@yukawa.ucsd.edu;
 Phone:(619)534-5742; Fax:(619)534-0173.}}}
\bi
\ct {\it Department of  Physics, University of California, San Diego }
\bi
\ct {\it 9500 Gilman Drive, La Jolla CA, 92093-0319. }
\vglue 2 cm
\ct {\bf ABSTRACT}
\bi

Calculations for inclusive $\psi$ production in $\Upsilon$ decay through the 
color-octet  mechanisms 
$b\bar b[^3S_1,\underline 1] \to c \bar c[^{2s+1}L_J,\underline 8] + g $ are
presented. 
It is shown that these ${\cal O}(\alpha^5_sv^4_c)$ contributions compete with 
other color-octet mechanisms considered in the literature. 
A critical numerical analysis of the color-octet contributions to $\Upsilon 
\to \psi + X$ shows that further work in this channel, both theoretical and 
experimental, is necessary in order to clearly understand the significance 
of the color-octet component of the $c \bar c$ component inside the $\psi$
 system. 
\bi
\noi
Pacs:13.25.Gv,12.38.Bx,14.40.Gx.

\vfill
\eject

\noi
{\bf I. INTRODUCTION}

 The unexpected  large rates for $\psi$ and $\psi^{\prime}$ prompt  production 
at the 
Fermilab Tevatron[1] gave rise to two interesting theoretical developments.
 On the one hand it was  realized that, at high energies, the most 
important contributions to $\psi$ production comes from processes  with a 
fragmentation 
interpretation[2]. On the other hand it was necessary to develop new 
ideas for heavy quarkonium 
 summarized in an effective theory called nonrelativistic QCD (NRQCD) [3]. 
This theory describes
 the interactions of non-relativistic quarks and is typically applied to
 $Q\bar Q$ bound states such as upsilon or $\psi$. The NRQCD Lagrangian is made
 precisely equivalent to full QCD through the addition of local interactions
 that systematically incorporate relativistic corrections through any given
 order in the heavy quark velocity $v$. The theory has a simultaneous 
expansion in $\alpha_s$ and in the velocity of the heavy quark. 
The size of a term in the NRQCD Lagrangian can be estimated using well defined
 velocity counting rules[4].

 Creation and annihilation of heavy quarkonium 
are naturally described within this theory which incorporates the 
factorization ideas. In fact, these processes, which necessarily 
occur at
 very small distances $\approx {1\over m_Q}$ ($m_Q$ denotes the heavy quark  
mass) as compared with the size of the meson
 $\approx {1\over m_Qv}$, are described by the short-distance
 coefficients 
entering in the relativistic corrections to the leading order NRQCD
 Lagrangian. The evolution of the quark pair  into physical quarkonium 
involves nonperturbative effects  which are encoded within matrix elements of
 higher order operators with a well defined hierarchy in the $v$ expansion.

As a consequence of NRQCD, the possibility arise for the formation of 
 physical quarkonium through higher Fock states. In particular, we have the
 possibility for the formation of physical quarkonium from a $Q\bar Q$ pair
 in a 
color-octet configuration. This is the most attractive explanation to
 the surplus $\psi$ and $\psi^{\prime}$ production at the Tevatron [5]. 

Unfortunately the NRQCD matrix elements can only  be computed from  lattice 
calculations which  are still in their infancy. The way people proceed is  to 
fit these parameters to some experimental data and use their values to make 
predictions for other processes where they are involved. Using the values for
 the NRQCD matrix elements as extracted from the fits to charmonium
 hadroproduction [6,17],
predictions have been made for color-octet signals in prompt quarkonium 
production in $ e^+ e^-$ colliders [7], $Z_0$ decays[8], photoproduction 
at fixed target experiments[9], hadroproduction experiments[10] and B-meson 
decays [11].

Inclusive $\psi$ production in $\Upsilon$ decay is a very good
 place to test the NRQCD ideas [12]. Unlike other $\psi$
 production processes , the short-distance coefficients highly suppress color
 singlet contributions to $\Upsilon \to \psi+ X$. In fact, color-singlet 
contributions start  at ${\cal O}(\alpha^6_s)$ through the 
$b\bar b[^3S_1,\underline 1] \to c\bar c[^3S_1,\underline 1] + gg$ and $b\bar 
b[^3S_1,\underline 1] \to
 c\bar c[^3S_1,\underline 1] + gggg$ mechanisms whereas 
color-octet contributions to the same process start at ${\cal O}(\alpha^4_s 
v^4_c)$
through the 
$b\bar b[^3S_1,\underline 1] \to c\bar c[^3S_1,\underline 8] + gg$  
mechanism. Hence, the short 
distance factors of the color-octet contributions are enhanced by
 factors of $ 1/\alpha^2_s$ as compared with the color-singlet ones. This 
enhancement 
easily overcomes the $ v^4_c$ suppression of the long
 distance factors. 

In principle we also have ${\cal{O}}(\alpha^4_sv^4_b)$ 
contributions from 
the $ b\bar b$ in a color-octet configuration. However, such contributions are 
naively suppressed by $(v_b/v_c)^4$ as compared with contributions from a
$ c\bar c$ 
in a color-octet configuration. Actual calculations reflect this naive 
counting suppression[12].

In reference[12] ${\cal{O}}(\alpha^4_sv^4_c)$ and leading electromagnetic  
contributions to 
$\Upsilon  \to \psi+ X$ were calculated within NRQCD. The most important 
mechanism turned out to be: $ b\bar b[^3S_1,\underline 1] \to c\bar c[^3S_1,
\underline 8] + gg$,
which gives a 
$BR\simeq  2.5 \times 10^{-4} $ when the value $\alpha_s(2m_c)=0.253 $ is
 used in the numerical calculations as suggested by the fragmentation 
limit for the $\Upsilon  \to \psi+ X$ decay width normalized to the 
$\Upsilon \to ggg $ decay width. 
 The next largest contribution  is the 
indirect production through $\chi_{cJ}$ decay considered by Trottier[13]. When
 the values for matrix elements extracted from the Tevatron
 data[6] are used, this contribution gives  $\sum_{J} BR(\Upsilon \to 
\chi_{cJ} 
+ X \to \psi + X) \simeq 5.7 \times 10 ^ {-5}$. There also exist  a  
comparable indirect contribution from $\psi^{\prime}$ having a branching 
ratio $\simeq 5\times10^{-5}$[12].
 Adding up all these and other even smaller  contributions a $BR \simeq 4
\times10^{-4}$ is obtained[12]. This result is within $2\sigma$ of the CLEO 
data:
$BR_{exp}=1.1\pm 0.4\times 10^{-3}$ [14] which suggest other  contributions 
need to be computed in order to bring theoretical predictions closer to 
experimental data.

Below I compute the ${\cal O}(\alpha^5_s v^4_c)$ one-loop  contributions to 
inclusive $\psi$ production in $\Upsilon$ decay
through the color-octet mechanisms $b\bar b[^3S_1,\underline 1]\to 
c\bar c[^{2s+1}L_J,\underline 8] + g$.
The importance of these mechanisms can be easily understood from the 
calculation for the electromagnetic contribution to the same processes[12]. 
This is an $ {\cal O}(\alpha^2_{em} \alpha_s v^4_c)$ contribution and 
gives a $ BR\simeq 1.6 \times 10^{-5}$. An ${\cal O}(\alpha^5_s)$ 
contribution
 should give roughly a
$ BR\simeq  10^{-4}$ or even larger if we consider the 
$e^2_be^2_c \simeq 5\times 10^{-3}$ factor
 coming from the quark  electromagnetic couplings. 
 The color-octet processes studied in this paper are also interesting due to 
the distinctive  signature that the $\psi$ has a sharply peaked energy 
distribution described by a delta function with a peak at $5.2$ GeV and 
is recoiled by a gluon jet. The delta function must be understood in the 
sense of NRQCD calculations, i. e., as a narrow distribution whose width is  
of order $m_cv^2_c$ in the $\psi$ rest frame. When boosted to the $\Upsilon$
 rest frame the width decreases by a factor ${m_{\psi}\over m_{\Upsilon}} 
\simeq 0.3 $, therefore giving 
narrow distribution peaked at $5.2$ GeV whose width is $\simeq 150 MeV$.

\noi
{\bf II. THE $b\bar b[^3S_1] \to c\bar c[^{2s+1}L_J,\underline 8] +  g$ 
MECHANISMS}

The factorization formalism developed in Ref. [3] can be easily generalized to
the case of inclusive charmonium production from bottomonium decay[13]. For the
 process under consideration in this work we have the factorization formula
$$
d\Gamma(\Upsilon \to \psi + X) = \sum_{m,n} d\Gamma(m,n) 
\bigl<\Upsilon|O(m)|\Upsilon\bigr>\bigl<O^{\psi}(n)\bigr>,\eqno(1)
$$
\noi
where $d\Gamma(m,n)$ is the short-distance factor for a $b\bar b$ pair in the
 state $m$ to decay into a $c\bar c$ in the state $n$ plus anything. The 
subscript
 $m, n$ denote collectively the color  and angular momentum quantum numbers of 
the heavy quark pairs. Contributions that are sensitive to the quarkonium
 scales (Bohr radius or larger) and to $\Lambda_{QCD}$  can be absorbed into
 the 
NRQCD matrix elements $\bigl<\Upsilon|O(m)|\Upsilon\bigr>, \bigl<O^{\psi}(n)
\bigr>$ . The relative importance
 of the
 various terms in the double factorization formula (1) depend on the order
 of $v$ in the NRQCD matrix elements and on the order of $\alpha_s$  in the 
short-distance factors. 

The Feynman diagrams for the ${\cal O}(\alpha_s^5v^4)$ contributions to 
the inclusive $\psi$ production in upsilon decay coming from the color-octet 
mechanisms   $b \bar b[^3S_1,\underline 1] \to c\bar c[^{2s+1}L_J,
\underline 8] + g$ are 
shown in 
Fig. 1.
Calculations for these processes are very similar to the calculations for
the radiative decays of quarkonium:
$Q\bar Q[^3S_1,\underline 1] \to q\bar q[^{2s+1}L_J,\underline 1] + 
 \gamma$ studied in reference [15].

Using the standard techniques for heavy quarkonium calculations [6,16,21] the 
following amplitude is obtained for  a $b \bar b$ pair annihilating
into three off-shell gluons: 
$$
{\cal M}[b \bar b[^3S_1] \to g_1(a) g_2(b) g_3(c)] = T_{abc} A, \eqno(2) 
$$
\noi
where $T_{abc}= {d_{abc}\over 4\sqrt{N_c}}$ is the color factor
$$A= ig^3_s\sqrt{{32\over \pi}}m_b{1\over  k_1\cdot(k_2+k_3) 
k_2\cdot(k_3+k_1)k_3\cdot(k_1+k_2)} LI$$
\noi
$ LI$ stand for the Lorentz invariant structure
$$
\eqalign{LI = 
 &\epsilon_1\cdot\epsilon_2 \bigl( k_1\cdot k_3
\epsilon_3\cdot k_2 \epsilon\cdot k_1 + k_2 \cdot k_3 \epsilon_3 \cdot k_1
\epsilon \cdot k_2 + k_1 \cdot k_3 k_2 \cdot k_3 \epsilon \cdot \epsilon_3
\big) \cr
&+\epsilon \cdot \epsilon_3 \big( k_1 \cdot k_2 \epsilon_1 \cdot k_3 
\epsilon_2 \cdot k_3 - k_1 \cdot k_3 \epsilon_1 \cdot k_2 \epsilon_2 
\cdot k_3 - k_2 \cdot k_3 \epsilon_2 \cdot k_1 \epsilon_1 \cdot k_3 \big) \cr
&+(\epsilon_1,k_1 \leftrightarrow\epsilon_3,k_3) + (\epsilon_2,k_2 
\leftrightarrow\epsilon_3,k_3); }\eqno(3) 
$$
\noi
here, $\epsilon$ denotes the $b\bar b[^3S_1]$ polarization vector and 
$\epsilon_i$
($k_i$) stand for the polarization vector (momentum) of $g_i$.

The amplitudes quoted in this paper are obtained by projecting the free quarks
amplitudes over definite  angular momentum ( $^{2s+1}L_J$) and color 
$(\underline 1,\underline 8)$ quantum numbers in the usual way [6,16].

The two gluon fusion into a $c\bar c[^{2s+1}L_J,\underline 8]$  processes are 
described by

$${\cal M}[g_1(a) g_2(b) \to ^1S_0,8c] = {d^{abc}\over 2\sqrt{2}}
g^2_s \sqrt{{2 \over \pi}}{1\over k_1\cdot k_2}\varepsilon (
\epsilon_1,k_1,\epsilon_2,k_2), \eqno(4)$$
$${\cal M}[g_1(a) g_2(b) \to ^3P_J,8c] = {d^{abc}\over 2\sqrt{2}}
g^2_s {4\over\sqrt{m^3_c}}\bigl({1\over k_1\cdot k_2}\bigr)^2
A_J,\eqno(5)$$
where
$$
\eqalign{A_0=\sqrt{1\over 6}\bigl[(k_1\cdot k_2 &+ 4m^2_c) (\epsilon_1\cdot
\epsilon_2 k_1\cdot k_2 - \epsilon_1\cdot k_2 \epsilon_2 \cdot k_1 ) + 
k^2_1 \epsilon_1 \cdot k_2 \epsilon_2 \cdot k_2 \cr
&+ k^2_2 \epsilon_1\cdot k_1 \epsilon_2 \cdot k_1 -
k^2_1k^2_2 \epsilon_1\cdot\epsilon_2  
- \epsilon_1\cdot k_1 \epsilon_2 
\cdot k_2 k_1\cdot k_2 \bigr]}$$
$$
A_1=m_c \bigl[k^2_1 \varepsilon (e^*,\epsilon_1,\epsilon_2,k_2) + 
\epsilon_1\cdot k_1 \varepsilon (e^*,\epsilon_2,k_1,k_2) + \{ k_1,\epsilon_1\}
\leftrightarrow \{k_2,\epsilon_2\} \bigr] $$
$$
A_2= \sqrt{8} m^2_c \bigl[ k_1\cdot k_2 \epsilon_{1\mu}
\epsilon_{2\nu} +
 k_{2\mu}k_{1\nu}\epsilon_1\cdot\epsilon_2 - k_{1\mu}\epsilon_{2\nu} 
\epsilon_1\cdot k_2 - k_{2\mu} \epsilon_{1\nu} \epsilon_2\cdot k_1 \bigr] 
e^{*\mu\nu},$$
\noi
here, $e^*(e^{*\mu\nu})$ denote the polarization vector (tensor) for the 
outgoing spin-1 (2) bound state and $\varepsilon$ stands for the Levi-Civita 
tensor.

Using the helicity projector techniques described in reference[15], 
the helicity 
amplitudes can be readily calculated for the different processes. A $1/2$ 
factor is inserted in the amplitudes for the full processes in order to 
avoid double counting of the one-loop diagrams as the amplitudes in 
equations(2),(4)and (5) are symmetric in the two gluons 
involved in the loop.  Performing the loop integration and matching the 
theories by the procedure described in reference [6] (alternatively the cross 
section for free quarks annihilation can be computed and matched with the 
NRQCD results by expanding in the quarks relative momentum and identifying 
the different terms by a procedure described in
 reference[21]) the following 
results are obtained for the short distance coefficients:

$$\Gamma[b \bar b[^3S_1,\underline 1] \to c\bar c[^1S_0,\underline 8], g] = 
Fr(1-r) |\hat H^s (r)|^2 \eqno(6)$$
$$\Gamma[b \bar b[^3S_1,\underline 1] \to c\bar c[^3P_J,\underline 8], g]=
{2F\over (2J+1)m^2_c}
r(1-r)\sum_{i=0}^J |\hat H^{(J)}_i(r)|^2 \eqno(7)$$ 
where
$$F={5 \over 32}{5\over 486} {\alpha^5_s \bigl<\Upsilon|O_1(^3S_1)|\Upsilon 
\bigr>
\over m^2_b m^3_c};\qquad  r= (m_c/m_b)^2 .$$

\noi
The explicit formulae for the helicity functions $\hat H^{(J)}_i$ are rather
 length
 and are deferred to the appendix.

The decay width for $\Upsilon \to \psi +  X$ is obtained by adding up the 
short-distance coefficients listed above 
multiplied by $\bigl<\Upsilon|O_1(^3S_1)|\Upsilon\bigr>$
 and their respective $\bigl<O^{\psi}_8(^{2s+1}L_J)\bigr>$ long-distance 
matrix elements:

$$\Gamma(\Upsilon \to c\bar c [^{2s+1}L_J,\underline 8 ]g \to \psi + X) = 
{5\alpha_s^5\over 486m^3_cm^2_b}
\bigl<\Upsilon|O_1(^3S_1)|\Upsilon\bigr>B \eqno(8)$$
\noi
where
$$B={5 \over 32}\bigl[f^s\bigl<O^{\psi}_8(^1S_0)\bigr> + \sum_{J=0}^2 
 f^{(J)}{\bigl<O^{\psi}_8(^3P_J)\bigr>
\over (2J+1)m^2_c}\bigr]. \eqno(9)$$

\noi
The $f$ factors appearing in equation(9) are obtained by evaluating 
$\hat H^{(J)}_i$ at $r= (m_c/m_b)^2 = 0.118$ :
$$f^s= r(1-r)|\hat H^s(r)|^2=5.79, \qquad   \qquad  
f^{(0)}=2r(1-r)\hat H^{(0)}(r)=6.54, $$
$$f^{(1)}=2r(1-r)\hat H^{(1)}(r)=7.32, \qquad   \qquad  
f^{(2)}=2r(1-r)\hat H^{(2)}(r)=8.34, $$
\noi
where
$$\hat H^{(J)}(r)= \sum_{i} |\hat H^{(J)}_i(r)|^2.\eqno(10)$$
Using the heavy quark symmetry relation $\bigl<O^{\psi}_8(^3P_J)\bigr>=(2J+1)
\bigl<O^{\psi}_8(^3P_0)\bigr>$, equation (9) can
 be written in the compact form:

$$B={5\over 32}\bigl[f^s\bigl<O^{\psi}_8(^1S_0)\bigr>
+f^p{\bigl<O^{\psi}_8(^3P_0)\bigr>
\over m^2_c}\bigr] \eqno(11)$$

\noi
where $$f^p = \sum_{J=0}^2 f^{(J)}=22.20 .$$ 

For numerical analysis purposes I list the decay width for inclusive $\psi$ 
production in $\Upsilon$  decay through the main
mechanism studied in reference [12]
$$ \Gamma(\Upsilon \to c\bar c[^3S_1,\underline 8 ]gg\to\psi + X)=
{5\pi\alpha_s^4\over 486m^3_cm^2_b}\bigl<\Upsilon|O_1(^3S_1)|\Upsilon\bigr>
\bigl<O^{\psi}_8(^3S_1)\bigr>\times (0.571). \eqno(12)$$

\vfill
\eject
\noi
{\bf 3. NUMERICAL ANALYSIS}

The decay width for $\Upsilon \to \psi + X $ in equations (8) and (12) 
depends on
 $\alpha_s$  and the NRQCD matrix elements 
$\bigl<\Upsilon|O_1(^3S_1)|\Upsilon\bigr>$ , 
$ \bigl<O^{\psi}_8(^3P_0)\bigr>$ , $\bigl<O^{\psi}_8(^3S_1)\bigr>$ and 
$\bigl<O^{\psi}_8(^1S_0)\bigr>$. In
 particular it is very sensitive to the chosen values for $\alpha_s$ ,
$\bigl<O^{\psi}_8(^3P_0)\bigr>$ and $\bigl<O^{\psi}_8(^3S_1)\bigr>$. 

\noi
The values for the color-octet matrix elements have been mainly extracted 
from hadroproduction and photoproduction data [6,11,17,18], these values being 
largely affected by NLO and perhaps by higher-twist corrections [17,19,20].
For the color-octet matrix element $\bigl<O^{\psi}_8(^3S_1)\bigr>$ a fit to 
CDF  data gives [6]
 $\bigl<O^{\psi}_8(^3S_1)\bigr> \in 
 [0.0045, 0.0087] GeV^3$ . The analysis of the same data including 
higher order QCD effects such as initial state radiation of gluons by the 
interacting partons[19] gives somewhat smaller values: 
$\bigl<O^{\psi}_8(^3S_1)\bigr> \in [0.0033, 0.0046] GeV^3$.
As to $\bigl<O^{\psi}_8(^3P_0)\bigr>$ and
 $\bigl<O^{\psi}_8(^1S_0)\bigr>$, different combinations of these matrix 
elements have been extracted from different hadronic
processes [6,11,17,18,19]. The values extracted for the same 
combination entering in different processes differ roughly by a factor of 2. 
In a recent analysis[20] which summarizes fixed target hadroproduction 
data[17], Tevatron data[6,19] and  photoproduction data[18], the following 
range was obtained for the combination of these matrix elements entering in
the inclusive $\psi$ production at Tevatron :
$\bigl<O^{\psi}_8(^1S_0)\bigr> + 3  \bigl<O^{\psi}_8(^3P_0)\bigr>/m^2_c  \in 
[0.01,0.06]GeV^3$.

On the other hand, $\psi$ production in $e^+ e^-$ annihilation was recently
 used to extract the same combination of matrix elements [22]. This process 
is not sensitive
to $\bigl<O^{\psi}_8(^1S_0)\bigr>$ but is highly sensitive to 
$\bigl<O^{\psi}_8(^3P_0)\bigr>$. 
The values obtained for 
$\bigl<O^{\psi}_8(^3P_0)\bigr>$ are very stable under changes in the input 
values for 
$\alpha_s$, charm quark mass and the color-singlet matrix element 
$\bigl<O^{\psi}_1(^3S_1)\bigr>$ entering in the fit,
 thus allowing us to strongly constrain the values for this matrix element:
  $\bigl<O^{\psi}_8(^3P_0)\bigr>/m^2_c \in [0.72, 0.76]\times 10^{-2} GeV^3$. 

In summary the only color-octet matrix element which seems to be firmly 
established
is $\bigl<O^{\psi}_8(^3P_0)\bigr>$ and Eqs.(8,12) should be analyzed as a 
function of the remaining color-octet matrix elements.

The values $\alpha_s(2m_c)=0.253, 
\bigl<O^{\psi}_8(^3S_1)\bigr>=0.014 GeV^3 , 
\bigl<\Upsilon|O_1(^3S_1)|\Upsilon\bigr>=2.3 GeV^3$ were used in [12]
in the numerical evaluation of equation (12). The value for
$\bigl<\Upsilon|O_1(^3S_1)|\Upsilon\bigr>$ is 
extracted from the Upsilon 
leptonic width and is very similar to potential model calculations, thus 
reflecting the suitability of the the non-relativistic description for 
bottomonium system.
The value for $\alpha_s$  is suggested by the fragmentation limit of 
equation (12) 
normalized to $\Gamma( \Upsilon \to ggg)$ which gives three times the 
fragmentation probability for a gluon fragmenting into a $\psi$, 
$P_{g \to \psi}$ [12,5]. However, fragmentation is not a good approximation 
for the process under consideration and the value used for 
$\bigl<O^{\psi}_8(^3S_1)\bigr>$ seems to be too large on the light of 
updated data[6,17,20]. 

A direct 
evaluation of equation (12) was not performed in reference[12]. Instead, 
authors evaluated this equation normalized to $\Gamma(\Upsilon \to ggg)$ and 
multiplied by $BR(\Upsilon \to ggg)$ which was assumed to be 
$\simeq BR(\Upsilon \to$
light hadrons$)= 0.92$. This procedure reduces the direct evaluation of 
Eq.(12) by a factor of $1\over 2.5$, which is compensated by the 
large value used for $\bigl<O^{\psi}_8(^3S_1)\bigr>$. 
A direct evaluation of Eq.(12) using $\alpha_s(2m_c)$ and the central  value 
 $\bigl<O^{\psi}_8(^3S_1)\bigr> \simeq 0.0066 GeV^3$ obtained in reference
 [6] gives a branching ratio $\simeq 2.6 \times 10^{-4}$. This $BR$ is 
reduced by a factor
$\approx {1\over 2}$ if the central value quoted in reference[20] for
$\bigl<O^{\psi}_8(^3S_1)\bigr>$ is used in 
the numerical evaluations.

Using the values
${\bigl<O^{\psi}_8(^3P_0)\bigr>\over m^2_c}= 0.0074 GeV^3 $ and 
$\bigl<O^{\psi}_8(^1S_0)\bigr>= 0.011 GeV^3$ as extracted  from $\psi$ 
production in $e^+ e^-$ annihilation [22] and  $\alpha_s(2m_c)=0.253$  the 
contributions calculated in 
this work gives a branching ratio 
$BR\simeq 2.1\times 10^{-4}$.
The pseudoscalar process Eq.(6) accounts for $\approx 25 \%$ of 
the contributions in equation (8), thus the
 branching ratio is less sensitive to $ \bigl<O^{\psi}_8(^1S_0)\bigr> $ than
to $\bigl<O^{\psi}_8(^3P_0)\bigr>$ . 
Adding up the contributions from equations(8) and (12) with 
other even smaller
contributions ( indirect psi production through $\chi_{cJ}$ and $\psi^{\prime}$
etc) calculated in Ref. [12] a total branching ratio 
$BR \simeq 6.2 \times 10^{-4}$ is obtained when $\alpha_s(2m_c)$ is used in 
the numerical calculations. However, this $BR$ is very sensitive to the chosen
value for $\alpha_s$.
\vfill
\eject

\noi
{\bf 4. CONCLUSIONS}

Summarizing, inclusive $\psi$ production in $\Upsilon$ decay through the 
color-octet mechanisms $b\bar b[^3S_1,\underline 1]\to c\bar c[^{2s+1}L_J,
\underline 8] + g $  is considered. A critical numerical analysis of these 
${\cal O}(\alpha^5_s v^4_c)$  contributions and other 
in the current literature is performed. The total branching ratio for 
$\Upsilon \to \psi + X$ and the size of the calculated contributions strongly 
depend on the values chosen for $\alpha_s$  and the color-octet matrix 
elements involved in the different mechanisms. Using $ \alpha_s(2m_c)$ as 
suggested by the fragmentation interpretation of the 
${\cal O}(\alpha_s v^4_c)$ term [12] and the central values 
 ${\bigl<O^{\psi}_8(^3P_0)\bigr>\over m^2_c}= 0.0074 GeV^3 $, 
$\bigl<O^{\psi}_8(^1S_0)\bigr>= 0.011 GeV^3$ as recently extracted from 
$e^+ e^-$ annihilation[22], the contributions considered in this paper give a
 $BR\simeq 2.1\times 10^{-4}$.
Adding up this branching ratio with the contributions calculated in 
reference [12] a total branching ratio $ BR \simeq 6.2 \times 10^{-4}$ is 
obtained in 
good agreement with the CLEO data $BR_{exp}= 1.1\pm 4 \times 10^{-3}$ and 
consistent with the ARGUS upper bound $BR_{exp}< 0.68\times 10^{-3}$ [14].
 
In the light of the available experimental information, the calculations 
presented in this paper indicates that color-octet 
mechanisms account for most of Psi production in Upsilon 
decay. However, further work, both theoretical and 
experimental, is necessary in order to have a clear idea on the role of the 
color-octet component of the $c\bar c$ pair inside the $\psi$ system.

From the 
experimental point of view, is necessary to remove the 
existing inconsistencies as the ARGUS collaboration results are not confirmed 
by the CLEO collaboration [14]. A measurement of the decay width and of the 
energy spectrum of the $\psi$ would be desirable as the later can 
discriminate between the different $\psi$ production mechanisms. In particular,
$\psi$'s  produced through the color-octet mechanisms considered in this 
paper have the distinctive signature that the
$\psi$'s  are sharply peaked in energy and are recoiled by a gluon jet. This 
is in contrast with other mechanisms where a spread distribution in energy is 
expected. 

From the theoretical point of view, the size of the calculated 
contributions strongly depend on the assumed value for $\alpha_s$. A numerical 
evaluation of equations(8,12) using $\alpha_s(2m_b)$ decrease the calculated 
$BR$ roughly by a factor of $1\over 5$. Hence, a calculation of the 
${\cal O}(\alpha^5_s v^4_c)$ color-octet mechanisms $b\bar b[^3S_1,\underline 
1] \to c\bar c[^{2S+1}L_J,\underline 8]ggg$ and the ${\cal O}(\alpha^6_s)$ 
color-singlet mechanisms 
$b\bar b[^3S_1,\underline 1] \to c\bar c[^3S_1, \underline 1]gg$ and 
$b\bar b[^3S_1,\underline 1] 
\to c\bar c[^3S_1,\underline 1]gggg$ is necessary in order draw definitive 
conclusions. The color 
singlet contributions have been ``crudely estimated'' to give a branching ratio
of a few $\times 10 ^{-4}$ [13].
\bi
\bi
\noi
{\bf Acknowledgments. }

I wish to acknowledge the hospitality of the high energy physics theory 
group at 
University of California, San Diego. 

\vfill
\eject
{\bf APPENDIX} 
$$\leftline{\indent$\displaystyle 
\eqalign {\hat H^s(r)=&{4\over x}\bigl[ 
(Di(2 x)-Di(0)-{x\over(1-2 x)}ln(2 x) - {1-x\over 2-x}\bigl(
 2 Di(x)-2 Di(0)+{1\over2}ln^2(1-x)\bigr) \cr
 &+  i 4\pi {1-x\over x(2-x)}ln(1-x)\bigr].}$}$$

$$\leftline{\indent$\displaystyle
\eqalign{\hat H^{(0)}_0(r)=&\sqrt{2\over3} \bigl[ {2-3 x\over x^2}  
 + \bigl(10 {1-x\over x^3} + 4 {1-2x\over x^2} ln(2) \bigr) ln(1-x) 
 - 3 {1-x\over x(2-x)}ln^2(1-x) \cr
&+  \bigl({8\over x^2} +2 {1-x\over x(1-2 x)} \bigr) ln(2 x) 
+ {8-6x+x^2-6x^3 \over x^3(2-x)} Di(0) - {4-5x+2x^2 \over x^3}
 Di(2 x) \cr
&- 4 {2-2 x-x^2 \over x^2(2-x)} Di(x) 
 + i 6 \pi {1-x\over x(2-x)} ln(1-x) \bigr].}$} $$

$$\leftline{\indent$\displaystyle
\eqalign{\hat H^{(1)}_0(r) =& {4\over x^2} \bigl[ {1\over x}(Di(0)-
Di(2 x)) - 6 (1-x) (Di(x)-Di(2 x)\cr
&-ln(2) ln(1-x)) + {2\over x} (1-x)
 (1-2 x) ln(1-x) + {2-8 x+7 x^2\over 1-2x} ln(2x)\bigr].}$}$$ 

$$\leftline{\indent $\displaystyle
 \eqalign{\hat H^{(1)}_1(r):=&{4 \sqrt{1-x}\over x^2} \bigl[ {1\over x}
 \bigl(Di(0)-Di(2 x)\bigr)-x-2 x \bigl(Di(x)-Di(2 x)
-ln(2) ln(1-x)\bigr) \cr
&+ {2-x-2 x^2 \over x} ln(1-x) +2 (1+x) ln(2 x) \bigr].}$} $$

$$\leftline{\indent$\displaystyle
 \eqalign{\hat H^{(2)}_0(r)=&{2 \sqrt{3}\over x^3} \big[ (6-5 x) x 
+ {2\over3} {6-19 x+18 x^2\over x} (1-x) ln(1-x) \cr
&-  {1\over 3} {10-12 x+5 x^2\over 2-x} \bigl(Di(0)-Di(2 x)\bigr) 
+ {2\over 3} {6-38 x+71 x^2-37 x^3\over 1-2x} ln(2 x) \cr 
&- 8 {(1-x)^2\over x^2 (2-x)}
\bigl( Di(2 x)-2 Di(x)-{1\over 2} ln^2(1-x)
+Di(0)+i \pi ln(1-x) \bigr) \cr 
&+ {4\over 3} {6-6 x-x^2\over x} 
\bigl(ln(2)-i {\pi\over 2}\bigr)\cr
&- {4\over 3 }\bigl(12-26 x+13 x^2\bigr) \bigl(Di(x)-Di(2 x)-ln(2) 
ln(1-x)\bigr) \bigr].}$} $$  
               
$$\leftline{\indent $\displaystyle
\eqalign{\hat H^{(2)}_1(r)=& {2\sqrt{1-x}\over x^3}\big[ -{1\over 3} 
(38-9 x) x - {2\over x} (4-13 x+16 x^2-4 x^3) ln(1-x) \cr
&+2 {x(1-x)\over2-x}
 \bigl(Di(0)-Di(2 x)\bigr) - {4\over 1-2 x} (2-11x+16x^2-4x^3) ln(2x) \cr
&+ 8 {(1-x)(2-2 x+x^2)\over x^2(2-x)} \bigl(Di(2 x)-2 Di(x)
-{1\over2}ln^2(1-x) + Di(0)+i \pi ln(1-x)\bigr)\cr  
&-{16\over 3} {3-3 x+x^2\over x} \bigl(ln(2)-i{\pi\over 2}\bigr) 
+ 4 (8-12x+3 x^2) \bigl(Di(x)-Di(2 x)-ln(2) ln(1-x)\bigr) \bigr].}$}$$

$$\leftline{\indent$\displaystyle
\eqalign{\hat H^{(2)}_2(r)=&\sqrt{2} {1-x\over x^3} \bigl[ {16 \over 3}x
 + {4\over x} (1-6 x+6 x^2) ln(1-x)-{2 \over 2-x} (5-6 x+2 x^2) 
\bigl(Di(0)-Di(2 x)\bigr) \cr
& -4 {2-4 x+6 x^2-4 x^3+x^4\over x^2 (2-x)} 
\bigl(Di(2 x)-2 Di(x)-{1\over 2}ln^2(1-x)+Di(0) + i\pi ln(1-x)\bigr)\cr 
&+{4\over 3} {6-6 x+11 x^2\over x} \bigl(ln(2)-i{\pi\over 2}\bigr) 
- 16 (1-x) \bigl(Di(x)-Di(2 x)-ln(2) ln(1-x)\bigr)\cr
&+ 4 (1-6 x) ln(2 x) \bigr].}$}$$
where $x=1-r$ and $Di$ denotes the Dilog function 
$Di(s)=\int_1^s{ln(t)\over 1-t}dt$.

\vfill
\eject

{\bf REFERENCES}

[1] CDF Collaboration F. Abe et.al. Phys. Rev. Lett 69,3704 (1992);
 71,2537 (1993);

\qquad 75, 1451 (1995). 

[2] E. Braaten and T.C. Yuan, Phys. Rev. Lett 71, 1673 (1993); Phys. Rev. D52,

\qquad 6627 (1995). 

[3] G.T.Bodwin, E. Braaten and G.P. Lepage, Phys. Rev. D51,1125 (1995); 
D55,5853

\qquad (1997); W.E.Cashwell and G.P. Lepage Phys. Lett. B167, 437 (1986); 
P. Labelle
 
\qquad hep-ph/9611313 (unpublished); A.V. Manohar Phys. Rev. D56, 230 (1997);
B.

\qquad Grinstein 
and I.Z. Rothstein hep-p/9703298 (unpublished); M.Luke and M.J. 

\qquad Savage hep-ph/9707313 (unpublished).

[4] G. Lepage et.al. Phys. Rev. D46 4052 (1992). 

[5] E. Braaten and S. Fleming Phys. Rev. Lett 74, 3327 (1995); M.Cacciari
et.al.

\qquad Phys. Lett. B536, 560 (1995). 

[6] P.Cho and A.K. Leivobich Phys. Rev. D53,6203 (1916).

[7] E. Brateen and Y.Q. Chen Phys. Rev. Lett 76, 730 (1996).

[8] K. Cheung, W.Y. Keung and T.C. Yuan, Phys. Rev. Lett.76, 877 (1996);

\qquad P. Cho, Phys. Lett B368, 171 (1996).

[9] J.P. Ma Nucl. Phys. B460, 109 (1996).

[10] K. Sridhar Phys. Rev. Lett 77 4880 (1996). 

[11] P. Ko, J. Lee and H.S. Song Phys. Rev. D53, 1409 (1996).

[12] K. Cheung, W.Y. Keung and T.C. Yuan Phys. Rev. D54, 929 (1996).

[13] H. Trottier Phys. Lett. B320, 145 (1994).

[14] CLEO Collaboration, R. Fulton et. al. Phys. Lett. B224, 445 (1989).

    \qquad ARGUS Collaboration, H. Albrecht et. al. Z Phys. C55, 25 (1992).

[15] J. G. Korner et. al. Nucl Phys. B229, 115 (1983).

[16]  J. H. Kuhn, J Kaplan and E.G.O. Safiani, Nucl. Phys. B157, 125 (1979);

\qquad B.Guberina et. al. Nucl. Phys. B174, 317 (1980).

[17] M. Beneke and I.Z. Rothstein Phys. Rev. D54, 2005 (1996); Phys. Lett. 

\qquad B372, 157 (1996).

[18] M.Cacciari and M. Kramer Phys. Rev. Lett 76, 4128 (1996); 

 \qquad J. Amundson, S. Fleming and I Maksymyk. hep-ph 9601298 (unpublished).

[19] M. A. Sanchis and B. Cano-Coloma hep-ph/9611264 (unpublished).

[20] S. Fleming et.al. Phys. Rev. D55, 4098 (1997).

[21] S. Fleming and M. Maksymyk. Phys. Rev. D54, 3608 (1996).

[22] F. Yuan C.F. Qiao and K.T. Chao, Phys. Rev. D56, 1663 (1997).

\vfill
\eject
{\bf FIGURE CAPTION}
\bi
\noi
Fig.1. One of the six diagrams for the short-distance processes
 $b\bar b[^3S_1,\underline 1]\to c\bar c[^{2s+1}L_J, \underline 8] + g. $

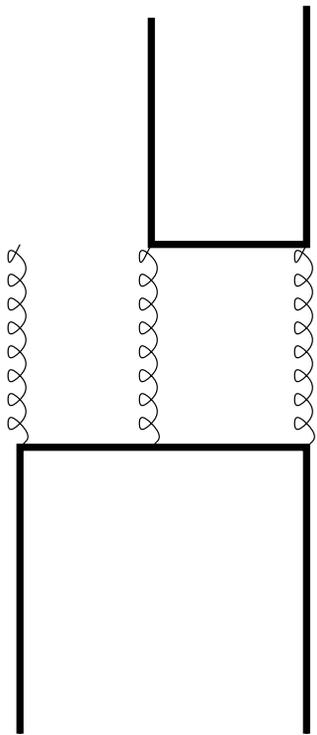
\end